# Description Languages for Consistency Management Scenarios Based on Examples from the Industry Automation Domain

Anthony Anjorin[a], Enes Yigitbas[a], Erhan Leblebici[b], Andy Schürr[b], Marius Lauder[c], and Martin Witte[d]

a   Paderborn University, Germany
b   Technische Universität Darmstadt, Germany
c   Continental Automotive GmbH, Germany
d   Siemens AG, Germany

**Abstract**   To cope with the increasing complexity of developing and maintaining modern (software) systems, multiple abstractions (models) of the same system can be established and used to allow different domain experts to collaborate and contribute their respective expertise. This divide-and-conquer, model-based approach requires, however, support for a *concurrent engineering process*, i.e., providing a means of checking, restoring, and ensuring the consistency of all involved and concurrently maintained models. The task of providing such support is often referred to as *consistency management*.

Although there exist various approaches to consistency management and numerous (industrial) case studies described in the literature on *bidirectional transformations* (bx), there is currently no *uniform* description of diverse but related industrial applications of model synchronisation and other forms of consistency management. This makes it challenging to detect similarities and differences related to requirements, constraints, applied techniques and tools. It is thus difficult to compare and transfer knowledge gained from (successful) projects to other bx approaches or even other bx tools for the same general approach.

In this paper, therefore, we propose a description language for envisioned scenarios in the problem domain of consistency management, as well as a complementary description language for solution strategies in terms of *method fragments* and *method patterns* in the solution domain of Model-Driven Engineering (MDE). Our work is inspired by previous research in the bx and MDE communities, and is also based on our collective experience from over ten years of investigating a series of application scenarios in the industry automation section together with Siemens AG as an industrial partner.

We use our proposed description languages to discuss a series of application scenarios that are diverse but all require varying forms of support for consistency management. By using a common notation and focusing only on aspects directly related to consistency management, we are able to abstract from project-specific details and uniformly describe how consistency management is required and can be currently supported in the industry automation sector. Based on this formal and macroscopic view of the projects, we provide a systematic discussion of our experience and results applying *Triple Graph Grammars* (TGGs) as a concrete bx approach in the industry automation domain.

# The Art, Science, and Engineering of Programming

**Description Languages for Consistency Management Scenarios**

# 1 Introduction and Motivation

The development and maintenance of increasingly complex software systems often requires a suitable decomposition of the system into multiple abstractions (models), which can be concurrently maintained by experts in their respective domains. Such a concurrent engineering process can be well supported by model-based approaches and techniques, especially approaches to ensure that the consistency of all related models, representing views of the same system, can be maintained via suitable change propagation and synchronisation. Numerous, diverse approaches to consistency management exist and are currently being actively researched and compared in the *bidirectional transformations* (bx) community [8, 36]. Involved communities in bx include databases (view-update problem, schema evolution), software engineering (model synchronisation, conformance checking), and programming language development (bidirectionalisation, coupled transformations).

Especially in a Model-Driven Engineering (MDE) context, Triple Graph Grammars (TGGs) [35] have often been used for consistency maintenance in various industrial applications [4, 17, 18, 34]. While this is certainly encouraging, it is currently challenging to detect similarities and fundamental differences in the ways TGGs have been applied to handle consistency management. The situation is even more critical for case studies using other bx approaches. This lack of a uniform, high-level but still precise description of application scenarios prevents the reuse and transfer of developed strategies and lessons learnt in these diverse projects to new projects. Having a priori knowledge about how to apply bx technology is indeed crucial, as an a posteriori refactoring of, e.g., unidirectional model transformations to utilise a bx language, is typically infeasible in practice. This situation effectively hinders the application and impact of bx technology in practice.

Depicted in figure 1, our first contribution (presented in section 2), is to suggest a description language that can be used to provide a macroscopic view on projects with varying requirements for consistency management in the problem domain. Inspired by and extending Stevens' work on describing *networks* of bx [37] we propose to define a *transformation context* specifying model types and consistency relations, and a *transformation schema* representing a basic pattern that all *networks* of models must adhere to. Based on this, it is then possible to describe typical *resolution paths* for the application scenario that specify how consistency is to be restored. Using this solution independent and technology agnostic description language, we present four different application scenarios of consistency management in the industry automation domain, investigated over the span of almost ten years in an ongoing cooperation together with Siemens AG as our industrial partner. The first project (completed), *Concurrent Model-Driven Automation Engineering* (CMDAE) [30, 34], is in the domain of automation engineering, the second project (also completed), *Concurrent Manufacturing Engineering* (CME) [2], in the domain of manufacturing engineering, the third (ongoing as of 2017), *A Graph-Grammar-Based Approach to Bidirectional Traceability of Related Models* (GraTraM), in the domain of computer-aided engineering. The fourth project is a project proposal based on research in the quality assurance domain [38].

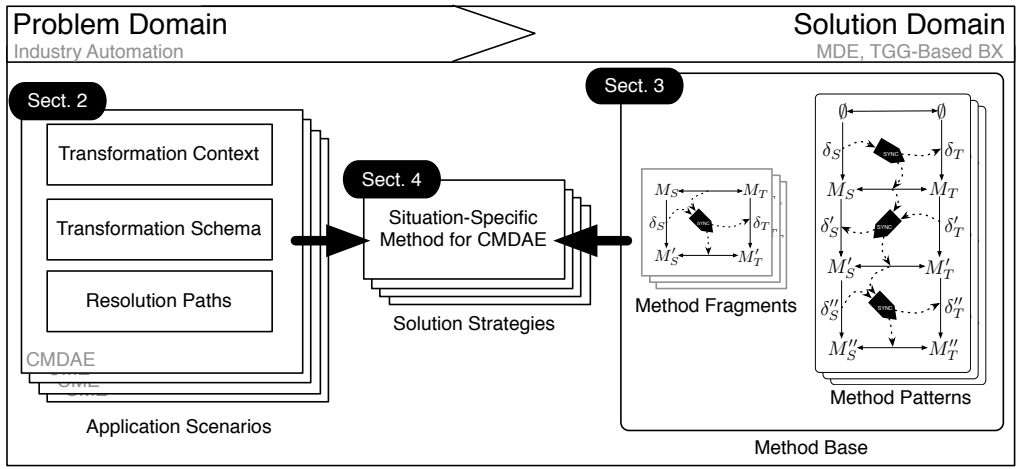

**Figure 1** Overview of the rest of the paper

Our second contribution (presented in section 3), is to propose a complementary description language for *method fragments* and *method patterns* used to realise MDE-based solution strategies. Our approach to describing solution strategies is based on *method engineering*, which is a discipline concerned with the systematic development and adaptation of *methods* [5]. A method is a description of how to systematically perform an endeavour and comprises a process together with relevant artefacts, roles, tools, and relationships between these elements on varying levels of granularity [19]. The main idea we apply (see figure 1), is to enable a modular construction of methods for consistency management scenarios by supplying a *method base* consisting of reusable atomic building blocks (method fragments), and construction guidelines on how to combine and assemble these fragments (method patterns) [19].

By providing a uniform description of all these diverse projects in the problem and solution domains, we are able to provide a discussion of project results in section 4 and identify current limitations and the primary potential of our applied TGG-based approach; by abstracting from project and tool-specific (technical) details we believe our contribution is comprehensible and accessible to other (bx) researchers, promoting a comparison and transfer of knowledge and insights to other approaches and domains.

In the rest of the paper, we give an overview of related work in section 5, and conclude in section 6 with a brief outlook on future work.

## 2 Describing Consistency Management Scenarios in the Problem Domain

In the following, we present our first project in section 2.1 and introduce our description language at the same time. This ensures that all formal definitions are directly followed by illustrative examples from the application scenario. Sections 2.2, 2.3, and 2.4 then present all other projects using the same description language. To keep this paper focussed, we are only able to provide a simplified view in each case, omitting many details and complexities discussed in the referenced literature for each project.

**Description Languages for Consistency Management Scenarios**

**2.1 Concurrent Model-Driven Automation Engineering (CMDAE)**

We commenced our research cooperation with Siemens AG by investigating how model-driven techniques can support the engineering of complex plants and machinery in the domain of automation engineering and manufacturing engineering (collectively referred to as the domain of industry automation in this paper).

The development of complex automation systems cuts across multiple engineering domains and requires the concurrent usage of different engineering tools. As the data exchange between such tools is often based on separate documents and meetings, a seamless integration of engineering models is required to increase the efficiency and effectiveness of the development process.

The case study investigated in the CMDAE research project was the development and maintenance of an automated storage and retrieval machine as part of a high-bay warehouse system [30, 34]. In this case study in the domain of automation engineering, the complete system specification is decomposed into three types of models: (i) *Location Oriented Structure (LOS)* models, representing the physical composition of a specific plant, (ii) *Hardware Configuration (HC)* models describing the logical interconnections between these components, and (iii) *Software Model*s (SM), the actual programming of automation devices in a language called *Statement List*.

The LOS is specified in IEC standard 81346. According to this standard, a plant is decomposed into different components, which may contain each other or be physically wired via ports. This information can be modelled using an engineering tool such as Comos ET.[1]

To model hardware centric information in the HC, the tool Simatic Step7[2] can be used, which also supports the development of Programmable Logic Controllers (PLCs). The HC contains abstract information regarding for example the type of a processor or communication module, while neglecting physical details such as the actual size or weight. Furthermore, these modules may communicate via logical connections, which do not necessarily reflect the actual physical wirings in the LOS.

The LOS, HC, and SM are *mutually dependent* on each other. For example, a change of an automation device in the HC due to performance requirements must be reflected in the bill of materials (a list of the raw materials and quantities of each required to manufacture an end product) of the electrical cabinet in the LOS. Analogously, a change of the device terminals in the LOS for optimised cabinet layout requires a change of the input/output signals used in control functions in the SM.

To formalise such a synchronisation scenario comprising multiple (types of) models and different consistency relations between them, we shall be using graphs and "arrows" between graphs based on the following standard definition (cf. e.g., [12]).

---

[1] http://w3.siemens.com/mcms/plant-engineering-software/en/Pages/Default.aspx?&L=1 (accessed 23.03.2018)
[2] http://w3.siemens.com/mcms/simatic-controller-software/en/step7/pages/default.aspx (accessed 23.03.2018)

**Definition 1 (Graphs and Graph Morphisms)**
*A graph $G = (E, V, src, trg)$ consists of a finite set $E$ of edges, a finite set $V$ of nodes, and two total functions $src : E \to V$ and $trg : E \to V$ that assign every edge its source and target nodes, respectively. Given graphs $G = (E, V, src, trg)$ and $G' = (E', V', src', trg')$, a (partial) graph morphism $f : G \to G'$ is a structure preserving pair of (partial) functions $(f_E : E \to E', f_V : V \to V')$, i.e., $f_E; src' = src; f_V$, and $f_E; trg = trg; f_V$, where ; is a composition operator such that $(f;g)(x) := g(f(x))$ for functions $f, g$ with suitable (co-)domains. Given graph morphisms $f = (f_E, f_V), g = (g_E, g_V)$, ; is a composition operator for graph morphisms such that $f;g := (f_E; g_E, f_V; g_V)$.*
*A graph monomorphism is a graph morphism $f = (f_E, f_V)$ with injective functions $f_E, f_V$.*

A *transformation context* is used to capture the types of models involved and the consistency relations between them. While this can be a hypergraph in general [37], our definition is simplified for the binary case as this suffices for all our projects:

**Definition 2 (Binary Transformation Context)**
*A binary transformation context $C$ is a graph whose nodes represent model types, and whose edges represent binary consistency relations between model types.*

Figure 2 depicts two diagrams: above a formal graph diagram with graphs as objects and graph morphisms as arrows, and below a diagram that introduces a suitable visual concrete syntax for our description language and exemplifies all defined concepts with the concrete CMDAE application scenario. In the graph diagram, monomorphisms such as $s$ are depicted as "harpoon" arrows. Let us concentrate for the moment on the transformation context C, which is depicted for the CMDAE application scenario to the far left of the diagram in concrete syntax. Note that there are dashed arrows connecting all abstract concepts (in this particular case C) to their concrete counterparts below.

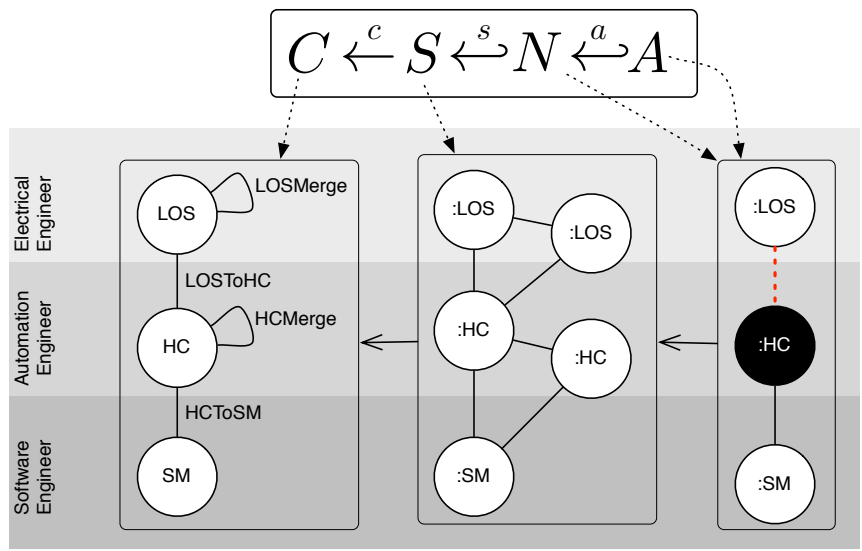

**Figure 2** Transformation context $C$ with schema $S$, and a concrete network $N$ with an authoritative sub-network $A$ for the CMDAE application scenario.

The CMDAE transformation context consists of the three model types already discussed previously (`LOS`, `HC`, `SM`) depicted as circles with labels. In addition, consistency relations `LOSToHC`, and `HCToSM` are used to pair consistent LOS and HC, and HC and SM models, respectively. Finally, two consistency relations `LOSMerge` and `HCMerge` are used to capture a notion of consistency between LOS and HC models of the same type. Note that while the edges in a transformation context are formally directed with a "source" and "target" model type, this is only used in practice to be able to sometimes refer to a "forward" and "backward" direction. We thus represent consistency relations as undirected labelled edges in our concrete syntax. As an additional (informal) extension to transformation contexts, we suggest to depict the different *roles* involved as horizontal swimlanes in the concrete syntax, containing all model types and models primarily maintained by the designated role: for the CMDAE transformation context, three different roles *Electrical Engineer*, *Automation Engineer* and *Software Engineer* are involved in the application scenario.

For a given transformation context $C$, we find it helpful to provide a *transformation schema S* used to fix a general "shape" or pattern that captures all possible concrete models and relations to other models. This is formalised with the following definition:

**Definition 3 (Transformation Schema)**
*A transformation schema for a given transformation context C is a graph S together with a graph morphism $c : S \to C$.*

Figure 2 depicts in the middle the schema for the CMDAE application scenario, connected to *S* in the formal graph diagram above. To denote the types of all models, i.e., the mapping to the underlying context *C*, UML-like labels of the form : *Type* are used e.g., : *LOS* as a model of type *LOS*. As the type (the corresponding consistency relation in the context) of all edges can be discerned uniquely in all diagrams, no explicit label is used. In the CMDAE scenario, the schema can be interpreted as the electrical and automation engineer each working with two branches: a `master` branch with LOS and HC models, and a `temporary` branch via which change requests can be made from other domains before being merged into the `master` branch. To discuss this further, we need the concept of a *transformation network* that captures a snapshot of all models and relations between models:

**Definition 4 (Consistent Transformation Network)**
*A transformation network for a given transformation schema S is a graph N together with a graph monomorphism $s : N \hookrightarrow S$. For a network N, an* authoritative *sub-network is a graph A with graph monomorphism $a : A \hookrightarrow N$ containing models that are* not *to be changed in the current network. An edge e in a transform network $N = (E, V, src, trg)$ is consistent if its source and target models $src(e), trg(e)$ are consistent with respect to the corresponding consistency relation $R = s_E; c_E(e)$ from the underlying transformation context. A transformation network N is consistent if all its edges are consistent.*

An exemplary transformation network for the CMDAE scenario (connected to *N* and *A* in the formal graph diagram) is depicted to the far right of figure 2. Every network must "comply" with its schema ($s : N \hookrightarrow S$ must exist) and is consequently "typed" according to the underlying context via $c : S \hookrightarrow C$ and $s; c : N \rightarrow C$. In this example, the *HC* is consistent with the *SM* but not with the *LOS*. Inconsistency between two models is indicated with a bold, dashed, red edge connecting the inconsistent models in the network. All other normal edges in the network are consistent. Models in the authoritative subnetwork are distinguished from other models by using a black background fill. In this example, the HC model in the network has a black fill and is thus in the authoritative sub-network (constituting *A* in the formal graph diagram). This means that it is not to be changed when restoring consistency in this network.

With our next and final definition, we can finally discuss typical synchronisation workflows that are to be supported in the CMDAE scenario.

**Definition 5 (Consistent Resolution Path)**
*A resolution path p is a sequence of partial graph morphisms $p_1; p_2; \ldots; p_k$ such that $p_i : N_{i-1} \rightarrow N_i$, all $N_i$ are transformation networks with a common schema S, and all $p_i$ are type preserving, i.e., $p_i; s_i = s_{i-1}$.*
*A resolution path of length k is* consistent *if the final network $N_k$ is consistent.*

Figure 3 depicts a typical resolution path for the CMDAE scenario. The sequence of networks is ordered from left to right, with white vertical lines between individual networks. Going from network $N_{i-1}$ to $N_i$, the underlying partial graph morphism $p_i$ is indicated visually using "version" numbers in the top-right corner of each node. Every model $m$ in $N_{i-1}$ for which there exists a model $m'$ of the same type and with the same version number in $N_i$, is "preserved" by $p_i$, i.e., $p_i$ is defined for $m$ and $p_i(m) = m'$. All other models in $N_{i-1}$ are "deleted" from the network, while all other models in $N_i$ that are not in the range of $p_i$ are newly "created". Note that all authoritative models in $N_{i-1}$ must be preserved in $N_i$ (but do not have to remain authoritative).

The workflow depicted in figure 3 starts with a trivial (empty) network $N_1$, which is assumed to be (trivially) consistent. The electrical engineer then commences the workflow by designing an initial version of the LOS, which can be optimised in several iterations until it is finally released in version 1 comprising the second network $N_2$. In $N_2$ the LOS is marked as being authoritative to make sure that it is not changed in the next step. Based on the initially created LOS, the automation engineer derives an authoritative HC that is consistent with the LOS in $N_3$. The process flows forward to the software engineer, who also derives an SM that is consistent with the current

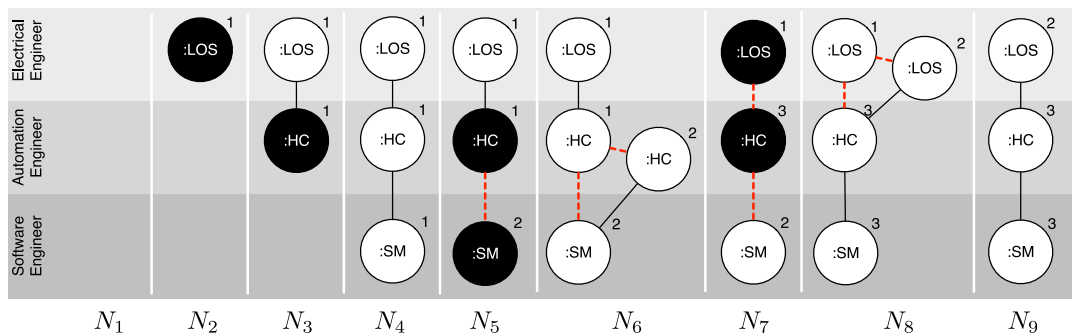

**Figure 3** A consistent resolution path for the CMDAE application scenario

HC in $N_4$. The software engineer, however, is not satisfied with this version and decides to make certain adjustments resulting in a version 2 of the SM in $N_5$, which is now inconsistent with the HC. In this network, however, both the HC and SM are authoritative, indicating a certain hierarchy between the HC and SM: instead of simply enforcing corresponding changes in the HC on the master branch, the software engineer is forced to "open a *temporary branch*" for the automation engineer by proposing a HC in version 2 (consistent with the SM but not with the master HC), representing the desired changes in the domain of the automation engineer.

The automation engineer now has to decide in $N_7$ how best to merge these changes and derive a consolidated HC in version 3, which is in general inconsistent with both the LOS and SM. To restore consistency in the network, the automation engineer is free to update the SM directly but, as the LOS is authoritative in $N_7$, must similarly open a *temporary branch* to the electrical engineer requesting a merge in $N_8$.

For this workflow, we now assume an optimistic setting where (i) the electrical engineer simply accepts all changes proposed by the automation engineer, and (ii) the software engineer is satisfied with all updates made by her colleagues. The network $N_9$ is thus in a consistent state.

## 2.2 Concurrent Manufacturing Engineering (CME)

In a second project CME [2], we investigated applying MDE technology to the domain of manufacturing engineering. Similar to projects in the automation engineering domain, manufacturing engineering projects are multi-disciplinary and require substantial coordination and information exchange among different engineering specialists, all concurrently using their own established tools, models and platforms.

Figure 4 depicts the transformation context and corresponding schema for the CME application scenario showing the different model types and roles involved: The task of the *product designer* is to design a *Computer Aided Design (CAD)* model providing a geometric representation of the product to be manufactured with a primary focus on aesthetics and functionality. The CAD model serves as requirements for a *manufacturing engineer*, who creates (and later updates) a corresponding *Computer Aided Manufacturing (CAM)* model specifying a sequence of basic manufacturing operations that must be executed to transition from given raw material to the end-product as required by the CAD model. As the CAD model might not be optimal for the manufacturing process, the manufacturing engineer and product designer typically work closely together, exchanging information and discussing possible CAM-friendly adjustments to the CAD model and their consequences.

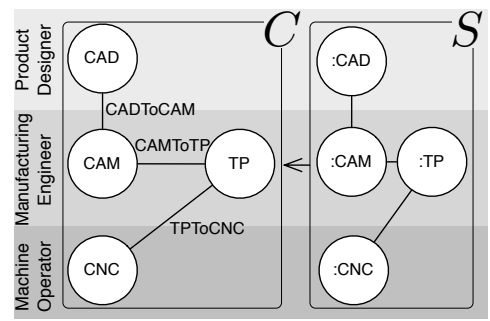

**Figure 4** Transformation context and schema for the CME project

The CAM model serves as input to a CAM tool that calculates and generates a corresponding *tool path* model TP, representing the exact movement of the tool tip of the involved manufacturing machine(s). Using this generated tool path model, the manufacturing engineer can execute the CAM model in a CAM simulator to further investigate and analyse the manufacturing process. Insights from the simulation are used in this step to further adjust and optimise the CAM model.

In a final step, the TP model is passed on to a *post processor*, a code generator that produces a *Computerised Numerical Control (CNC)* model, i.e., a program (code) that can be executed to drive CNC machines that perform the actual manufacturing process. In contrast to the CAM and TP models, the generated code is typically machine (family) specific as the exact kinematics (geometry of motion) of the machines is exploited to produce highly optimised code.

As a final role, a *machine operator* supervises the manufacturing process and can decide to intervene and correct or further optimise the CNC code. As an example, the machine operator could be more familiar with the exact manufacturing machine at hand and might realise that certain operations can be further accelerated. As the CNC development environment is, however, not yet fully integrated in typical CAM tools, such changes are often made directly to the generated CNC code and have to be later translated into corresponding adjustments to the CAM model. Investigating and supporting this final round-trip between adjusted CNC and CAM models was the main focus of the CME research project. For this application scenario, the schema comprises exactly one model of each type. Direct consistency relations are defined between CAD/CAM, CAM/TP, as well as TP/CNC models.

Figure 5 depicts an exemplary resolution path for the CME application scenario. The path starts with a consistent network $N_1$ comprising all models in version 1. The main scenario investigated in the CME project starts in $N_2$ with a change made by the machine operator to the CNC model. To avoid losing such changes when re-generating the CNC code, the new CNC model in version 2 is marked as being authoritative. It is in general inconsistent with the existing TP model. In $N_3$ the manufacturing engineer updates the TP model and can inspect any inconsistencies between the updated TP in version 2 and the CAM model. The manufacturing engineer decides exactly how to update the CAM model, perhaps accepting some implied changes and rejecting others. In $N_4$ the manufacturing engineer is satisfied with the new version 2 of the CAM model and has re-generated a consistent TP model. Both CAM and TP

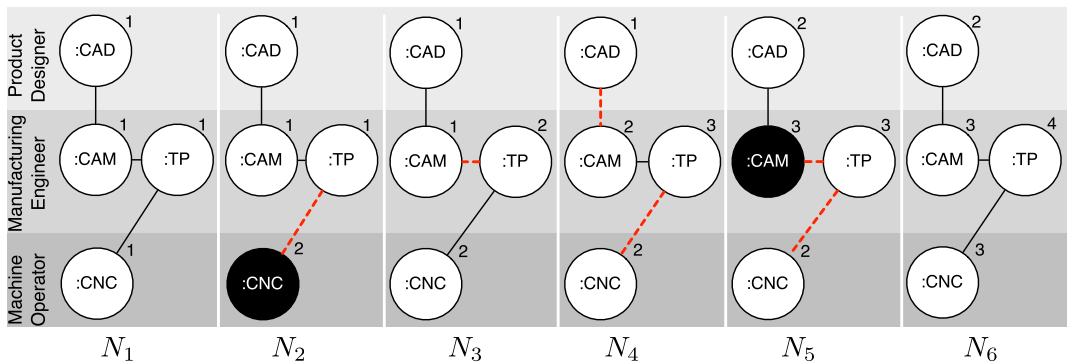

**Figure 5** A typical resolution path for the CME application scenario

**Description Languages for Consistency Management Scenarios**

are now inconsistent with the CAD and CNC models. In $N_5$ the product designer and manufacturing engineer have decided together how to transition to consistent states of both CAD and CAM models now in version 2 and 3, respectively. The CAM model is made authoritative to indicate that only the TP and CNC models are to be changed to restore consistency with the CAM model. In $N_6$ the TP is re-generated and from that a new version 3 of the CNC model is derived. The path is now consistent and we assume that all important changes made to the CNC code have been reviewed and integrated in all other models.

## 2.3 Graph Grammar-Based Traceability Management (GraTraM)

In a third ongoing research project (GraTraM)[3] we are currently investigating consistency management tasks in the domain of Computer-Aided Engineering (CAE).

As depicted in figure 6, the involved models in this domain are CAD models, providing a physical description of a system (for example an excavator in a current case study), and simulation models (Si), used to assess the behaviour of mechanical, hydraulic, and electrical components of the same system. Similar to models in the domains of automation and manufacturing engineering, CAE models are maintained by different domain experts with different engineering tools: CAD models with NX[4] used by CAD engineers, and simulation models with Amesim[5] used by simulation engineers. The consistency relations in this domain are between CAD and simulation models, as well as between different CAD models representing a similar "merge" as in the CMDAE application scenario.

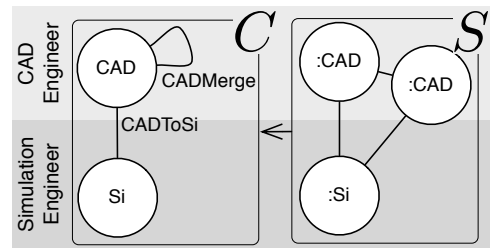

**Figure 6** Transformation context and schema for the GraTraM application scenario

Consistency management tasks in this application scenario typically involve geometric data and material properties of individual components. The shared information between CAD and simulation models, however, constitutes only a small subpart of these models. In practice, therefore, simulation models are typically *not* (partially) derived automatically from CAD models (or vice-versa). As depicted in figure 7, therefore, a typical GraTraM resolution path starts with first versions of both models that have been *concurrently* and largely independently developed (as opposed to a forward engineering step used to derive an initial version of one of the models). Both engineers have the primary task of specifying their own models at the beginning of the process and existing solutions are often re-used at this stage. In order to execute a simulation, both models must agree on geometric data and material properties of the system.

---

[3] Planned project end is February 2018.
[4] http://www.plm.automation.siemens.com/en/products/nx/ (accessed 23.03.2018)
[5] http://www.plm.automation.siemens.com/en/products/lms/imagine-lab/amesim/ (accessed 23.03.2018)

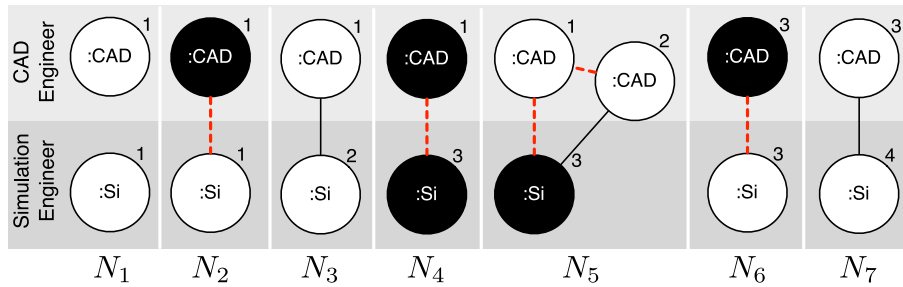

**Figure 7** A typical resolution path for the GraTraM application scenario

As depicted in $N_2$, the CAD model is authoritative for this step meaning that the simulation model must be adjusted to be consistent with the current CAD model. After restoring consistency and ensuring that the two models agree on their shared information in $N_3$, a simulation can now be performed. Based on simulation results, the simulation engineer now decides in $N_4$ to perform some optimisations by adjusting the simulation model until satisfactory results are produced. This leads to inconsistencies with the CAD model, which cannot be directly manipulated to restore consistency (both models are authoritative in $N_4$). The changes to the simulation model must, therefore, be reflected in a consistent CAD model (now in version 2) and passed to the CAD engineer, who decides how to incorporate and "merge" these changes into the previous master CAD model resulting in a version 3 of the CAD model in $N_6$. The CAD engineer can now restore consistency in $N_7$ by manipulating the simulation model.

## 2.4 Model-Based Transformation Testing (MBTT)

Our final project is concerned with the automated testing of a complex code generator (the System Under Test: SUT) used to transform CAM models to CNC code (as required for the CAD-CAM-CNC manufacturing engineering process chain discussed in section 2.2 for the CME project). This application scenario is thus in the intersection of the manufacturing engineering and quality assurance domains. The transformation context and schema for this scenario are depicted in figure 8. Relevant for this scenario is checking for the consistency of specified CAM and generated CNC models.

One might expect testing in this context to be a straightforward process of establishing an expected version of the CNC code and comparing this to the generated CNC code from each test run. The challenge here, however, is that the code generation process is highly non-deterministic. The reason for this is that the final result of a manufacturing (sub) process as specified by the CAM model has some degrees of freedom regarding the exact order in which atomic steps are to be executed by a CNC machine. As an example, consider drilling a countersunk hole for a screw, i.e., a hole with a recess for accommodating flathead screws to ensure that they are flush with the surface. The code to accomplish this task can instruct the drilling machine to first of all drill the recess, then the actual hole, or vice-versa. As long as the manufacturing task specified in the CAM model is accomplished, the generated code is "correct" and is expected to pass all tests. In a CAM model of realistic complexity, numerous such non-deterministic choices are combined, leading to a fairly large solution space of correct CNC programs; this makes testing such a code generator relatively challenging.

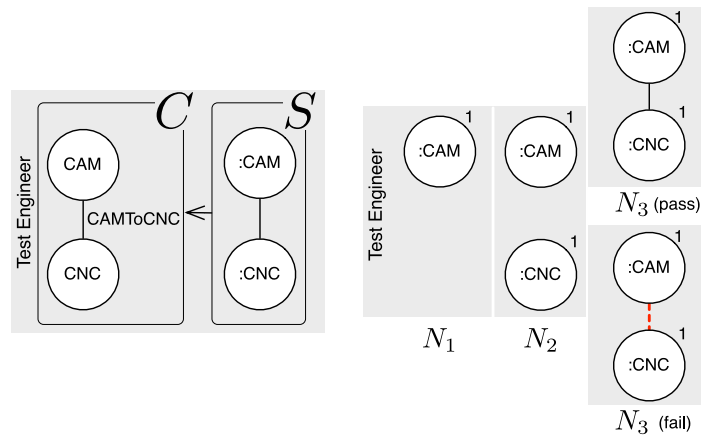

**Figure 8** MBTT context, schema, and resolution paths as test outcomes

The testing process can be viewed as checking for possible (consistent) resolution paths as depicted in figure 8. A *test engineer* has the initial task of specifying CAM models as test input data in $N_1$. The next task in $N_2$ is to generate CNC code as test output data using the post-processor (the SUT). A verdict is then determined using a suitable test oracle that decides if the CAM and CNC models are consistent or not. These two possibilities are depicted in figure 8 as $N_3(pass)$, i.e., the models are consistent and the test case passes, and $N_3(fail)$, i.e., the models are inconsistent and the test case fails. Note that there are no authoritative models in this resolution path as the goal is only to determine if the two models are consistent or not, and not to restore consistency.

## 3 Describing Consistency Management Scenarios in the Solution Domain

All application scenarios we have discussed so far have in common that they are inherently multi-disciplinary, requiring substantial cross-domain collaboration and data exchange between different tools and platforms. Multiple steps are required to (re-)establish the consistency of multiple models. While this can be achieved manually, some support for automation would be advantageous in all scenarios.

Even with a uniform, high-level and simplified description for the scenarios as we propose in section 2 in the problem domain, it is still non-trivial to establish correspondingly high-level and reusable *solution strategies* due to non-standard terminology, diagrams, and other methods of representation used in the solution domain. We argue that this is equally important as scenario-specific and often technical details can easily obfuscate steps in the process that could be addressed by applying the same techniques and technology.

The objective of this section is, therefore, to establish a common *method base* that can be used to provide a uniform description of high-level solution strategies for application scenarios. Such a method base consisting of reusable fragments and patterns can help to transfer experience and even implemented tools between projects. As our chosen solution domain is MDE, we propose to define a description *modelling language* for the solution domain in typical MDE style: Figure 9 depicts a *metamodel* to the left specifying the structure (abstract syntax) of valid models and some basic constraints that must not be violated. A proposed visual notation (concrete syntax) for the description language is depicted below the metamodel. According to Engels et al. [13], a *software engineering method* is used to systematise an endeavour by specifying the activities to perform, artefacts to create, roles to involve, tools to use, and relationships between all these concepts. A *method fragment* is a reusable atomic building block of a method [22] and can represent any constituent of the method, such as a tool to use or a role to involve. We propose in this paper to focus on just activities (Activity) and artefacts (Artefact).

Our proposed metamodel is thus a small subset taken from the metamodel suggested by Engels et al. [13] and could be extended as required to cover roles, tools, disciplines, domains, etc. As method fragments correspond in our context to atomic executable components that can be directly derived from a bx (in the following a TGG, a set of constraints, or a program in some other bx language) we introduce some additional subtypes of activities and artefacts. We focus on three different activities: consistency checking (CC), model generation (GEN), and synchronisation (SYNC). An activity can have input (in) and has output (out) artefacts. All activities are represented in the concrete syntax as black fat arrows with a corresponding label. Correspondences (Corrs) are artefacts representing a connection between a source (src) and a target (trg) model (Model), which are also artefacts. As in the problem domain, source and target can be viewed as arbitrary domain labels and enable a discourse about a "forward" or "backward" direction. Correspondences are denoted by horizontal double headed arrows. The concept of a delta (Delta) specialises correspondences and represents connections between models in the same domain, in this case typically changes applied to derive the target model from the source model. Deltas are denoted by vertical single headed arrows. Models can be empty as it is often useful to distinguish the special case of creating an artefact from scratch. Models are generally denoted with an "M" possibly with a subscript to indicate domains, while empty models are denoted by the empty set symbol ∅. We choose to denote the trivial "empty" correspondence between empty models explicitly as a horizontal double headed arrow to indicate that a pair of empty models is indeed consistent. This does not have to be the case in general.

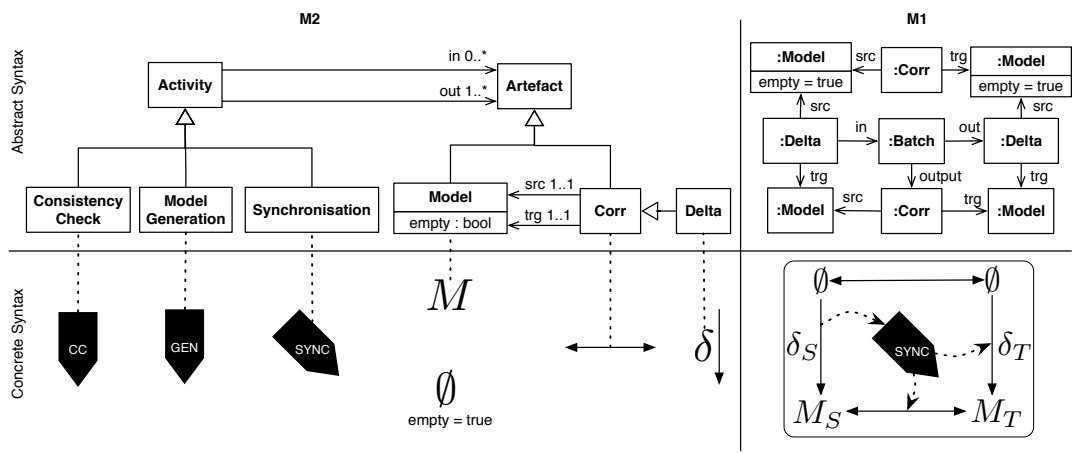

**Figure 9** Description language for solution domain

An exemplary instance of this metamodel, a method fragment, is depicted to the right of figure 9 in abstract syntax above and concrete syntax below. In concrete syntax, input and output artefacts are connected to the activity via dotted arrows.

## 3.1 Method Fragments for Consistency Management

Using this description language for method fragments, we can now present the eight method fragments that we have identified and used for establishing consistency management solutions in the industry automation domain and in other projects. The fragments are depicted in figure 10 organised in three top-level groups. Note that each fragment operates on concrete models, and their implementations are to be derived from a consistency description between two meta-models. Providing the consistency description itself, of course, relies on domain-specific knowledge capturing concepts and their intra-model as well as inter-model relationships, while realising these fragments requires bx expertise.

The top-left group GEN contains two fragments for generation: *initial* model generation to the left, and *incremental* model generation to the right. These two fragments imply that it is possible to derive a model generator from a bx that is able to generate consistent models from scratch as well as incrementally extend a consistent pair of models connected by correspondences (a consistent *triple* in the following).

The left-bottom group CC contains two fragments for consistency checking: to the left an initial check given two models, and to the right the incremental case for changes to previously consistent models. Note that the output of the consistency check is the correspondence between the two final models. For inconsistent input, we suggest that this check not only report that the input is inconsistent, but also identify a consistent subpart of the input that is consistent and for which correspondences can be created. We refer the interested reader to Leblebici et al. [32] for further details.

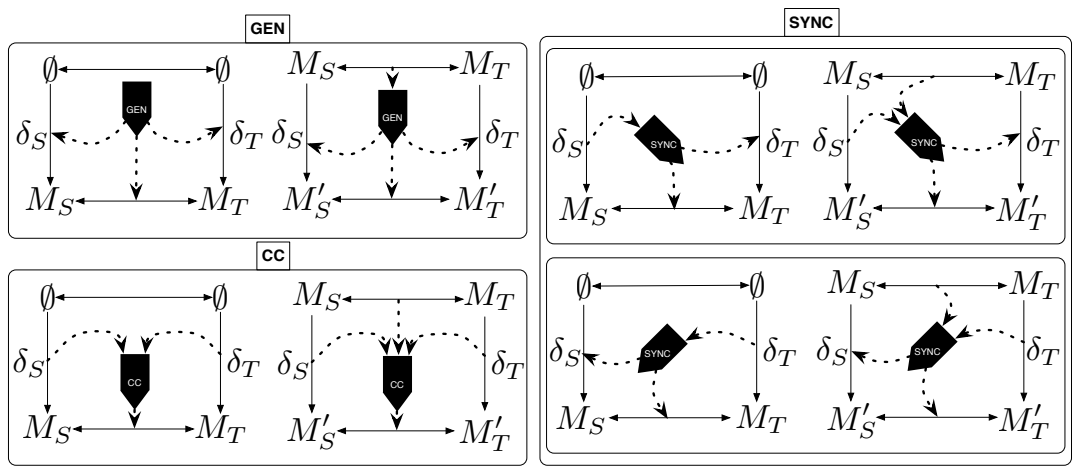

**Figure 10** Method fragments for consistency management

The third top-level group SYNC to the right of figure 10 contains a subgroup for *forward* synchronisation, and one for *backward* synchronisation. As for GEN and CC, each subgroup distinguishes an initial case for creating a new target/source model from scratch, and an incremental case for applying suitable changes to an existing target/source model.

The method fragments described so far are atomic steps that can be combined as necessary. In the next section we shall now combine them to present some high-level solution strategies as method patterns.

### 3.2 Method Patterns for Consistency Management

A *method pattern* provides construction guidelines for a method by specifying which method fragments to use and how to assemble them [20]. In this paper, we present every method pattern with an *Intent*, the concern or problem that it addresses, and a *Strategy*, the methodological solution it follows [20].

Our method patterns represent solution strategies that we have applied in multiple projects and examples, including the four scenarios discussed in this paper. Two patterns discussed in the following section concern the start of a synchronisation scenario, while a third pattern concerns how a bx can be used to test an independent implementation [26, 38].

#### 3.2.1 Initial Method Patterns: DirectedStart and ColdStart

Starting a consistency management process in general, and a synchronisation process in particular, is challenging. This is because the various tools pose different assumptions and requirements that must be fulfilled at the start of the process.

For model synchronisation, one of the assumptions is that we are either (i) using the initial method fragment and can establish an initial correspondence while creating the output model from scratch, or (ii) using the incremental fragment and already have a previous correspondence that serves as input for SYNC. The latter case is not very helpful when trying to initiate the synchronisation process. Depicted in figure 11 to the right, our first method pattern DirectedStart represents the easier case, when the application scenario allows starting with a "forward engineering" step, where one model $M_S$ is cre-ated and then forward synchronised to produce the other model $M_T$ from scratch, producing an initial correspondence between these two models in this step. After establishing a consistent triple in this manner, changes made to one model can then be propagated to the other model using the incremental fragments for forward/backward synchronisation as depicted in figure 11.

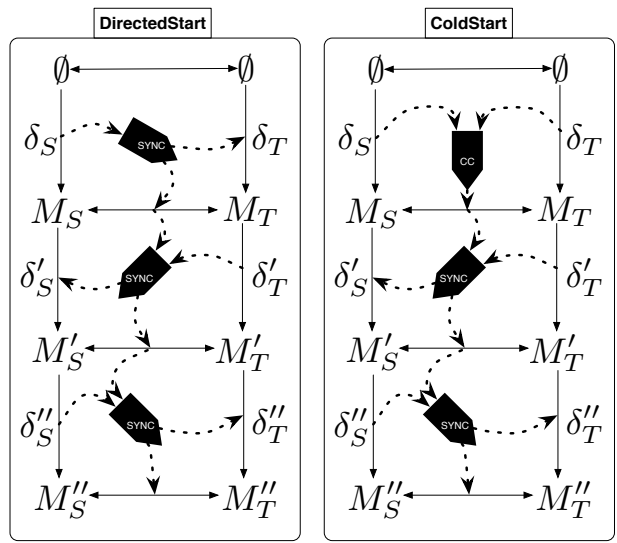

**Figure 11** Initial method patterns

The intent of `DirectedStart` is thus starting a synchronisation process, while its strategy is exploiting the fact that it is easy to establish an initial correspondence while creating the output model from scratch. This can also be typically accomplished in a scalable manner.

A second "initial" method pattern that also represents a strategy for starting a synchronisation process is `ColdStart`. As depicted in figure 11 to the right, the method pattern `ColdStart` represents the case when an application scenario forces a start with two existing (and possibly independently developed) models $M_S$ and $M_T$ and no correspondence between them yet. To be able to run `SYNC`, therefore, a correspondence must first be established using the initial `CC` fragment. While this can be handled [32], the situation is more challenging than `DirectedStart` due to a potentially very large solution space. Scalability is hard to achieve and a manual implementation (without a solver of some kind) is difficult. Realising early enough that a `ColdStart` is absolutely necessary for an application scenario is thus important.

The intent of `ColdStart` is the same as for `DirectedStart`, i.e., kickstarting a synchronisation process, but its strategy is to establish a correspondence between two unrelated models in an attempt to extend them to a consistent triple, with which incremental `SYNC` fragments can then be executed.

### 3.2.2 Method Pattern for Quality Assurance: GenerateAndCheck

For scalability reasons or due to other practical and technical (application-specific) requirements or constraints, a transformation or bx sometimes has to be implemented in a language that might not be particularly suitable for doing this.

In such cases, a high-level implementation with formal guarantees or improved readability (and thus trustworthiness) can still be used as an oracle for testing the low-level implementation [26, 38]. This idea is represented by the method pattern `GenerateAndCheck` depicted in figure 12.

To test the System Under Test (SUT), a test generator providing test input data and a test oracle deciding if the SUT behaves as expected are required. The initial and incremental `GEN` fragments can be used to create and extend consistent triples to be used as test input data for the SUT. To test a forward/backward transformation, the generated source/target model is passed to the SUT to create its output target/source model. The model created by the SUT can be tested by checking for correctness using initial and incremental `CC`, i.e., checking if the generated input model and the output model produced by the SUT are consistent. Note that as there might be multiple consistent $M_T$ for test input $M_S$, simply checking that $M_T = M_T^{SUT}$ is typically naïve, which is why `GenerateAndCheck` proposes to use the `CC` fragment for this.

The intent of `GenerateAndCheck` is to systematically test a bx implemented in a low-level, legacy, or otherwise unsuitable language. The employed strategy is to combine `GEN` and `CC` fragments as test generator and oracle, respectively.

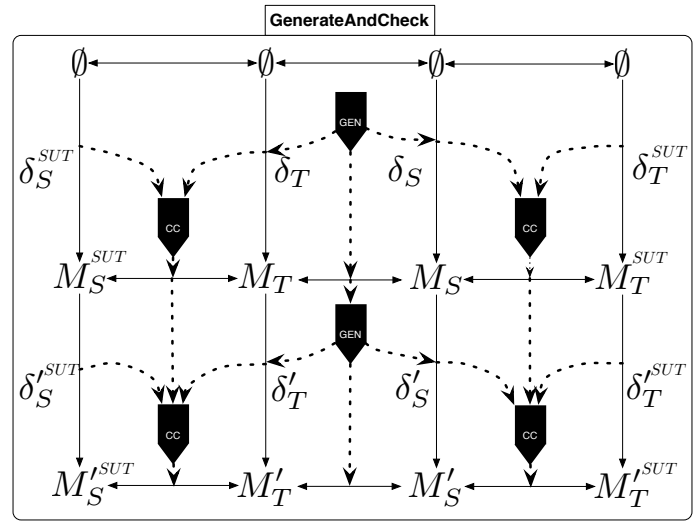

**Figure 12** Method pattern for model-based transformation testing

## 4 Discussion

In the following, we discuss several points based on applying our proposed fragments and patterns to the application scenarios presented in section 2: (i) What could be automated in the project, what not, and why not? (ii) What is the initial cost of using a bx language? (iii) How could primary resolution processes be started? (iv) Was a bx language such as TGGs really helpful and what are general indications for this? And finally, (v) How formal and on which level of abstraction should consistency management description languages be?

### 4.1 What can be currently automated (with or without a bx language)?

The languages we propose in this paper are currently descriptive rather than prescriptive. Our aim is to document past projects and implemented solutions in a high-level but still precise manner and thereby foster reuse of solution ideas (patterns) in future projects. Many actively developed bx tools [24, 27] support and focus on automating SYNC fragments, typically requiring one direction and automatically providing the other direction. Some bx tools [17, 26, 31], typically TGG-based, require a GEN fragment and automatically provide SYNC and in some cases even CC fragments. Other bx tools [6, 33] require a constraint-based definition of consistency and (ignoring scalability issues) are often able to provide all fragments. Using our languages, it is not yet possible to automatically transform a description of a scenario in the problem domain to a possible description in the solution domain. We also do not yet support generating a chain of consistency management actions from our patterns.

We now discuss the level of automation we were actually able to achieve for our industry automation scenarios: For the CMDAE scenario, we were able to specify the consistency relations LOSToHC and HCToSM to a certain extent (limited by the time available for the projects) using TGGs [30, 34]. Both initial and incremental forward

**Description Languages for Consistency Management Scenarios**

and backward SYNC fragments were useful for the projects. The consistency relations LOSMerge and HCMerge, both representing a merge of models of the same type, were considered out-of-scope and could only be performed manually.

Supporting such merge operations to restore consistency is challenging as it involves supporting both a conflict detection and a conflict resolution process. This can be represented as the model fragment INT depicted in figure 13, generalising all other model fragments. Deriving a scalable and "well-behaved" INT automatically from a bx is an open research question and a current focus of the bx community.

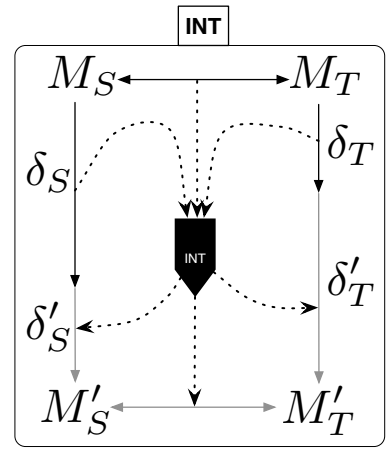

**Figure 13** Method fragment INT for supporting model integration

For the CME scenario, the consistency relation TPToCNC was the sole focus of the project [2] as all other consistency relations were already well-supported and integrated in a typical CAM/CAD tool. An observation is that such tools do *not* automate, e.g., the relation CAMToTP or CADToCAM, but instead support consistency restoration by providing, e.g., a visual diff of inconsistencies. Indeed it is questionable whether such tasks could ever be (fully) automated as they involve design decisions; in this case a compromise between aesthetics and pragmatics of manufacturing. For the GraTram scenario, the CADToSi consistency relation was investigated and specified using TGGs. Similar to the merge relations in CMDAE and CME, the CADMerge relation is manually checked and maintained; it is again arguable whether such a design process could ever be automated meaningfully.[6] Finally, the primary goal in the MBTT scenario was to automate consistency checking for the CAMToCNC relation.

To summarise, our observation is that synchronisation and consistency checking tasks can be automated if there is no conflict detection and resolution involved. This is, for example, the case when changes to only one of the models are to be propagated to all other models. Supporting conflict resolution is the focus of current research and, based on our projects, we believe that this must involve user interaction in cases where the conflicts imply design decisions that probably cannot be fully automated.

### 4.2 What is the (initial) cost of using a bx language?

As with any new (programming) language, an initial cost for using a bx language such as TGGs is certainly the effort required to learn the language and become productive in it. For TGGs this requires rule-based thinking in terms of graph patterns, and an acceptance/appreciation of an inherently visual formalism. Our experience in the industry automation domain indicates a promising acceptance from domain experts with experience using UML and other MDE technology, but this can be (very) different in other domains where other bx approaches might be more successful. With reasonable effort we believe domain experts can read and understand TGGs; specifying TGGs probably requires substantially more training effort.

---

[6] Possibly by applying machine learning algorithms which was out-of-scope for the project.

An integration with the existing tool chain is a very important requirement in the domain of industry automation as all tools and formats are fairly established and in many cases already standardised. This implies that *tool adapters* must be implemented to extract and inject the required data from the relevant tools.

In the case of the CMDAE scenario, we were able to use an XML-based exchange format for LOS, and text (code) for SM. We had to implement a (highly simplified) parser for SM, however, as we were unable to reuse the existing parser in the actual system. For the CME scenario a textual exchange format for TP could be used and CNC is already present as code. We had to again implement a (naïve) parser for a set of CNC examples as it was again out-of-scope to invoke and integrate standard parsers that were unfortunately tightly coupled with their tool environments. Implementing such a parser is difficult as CNC code can be (in general) interwoven with parts in assembler. For the GraTraM scenario establishing a tool adapter for the CAD tool was especially challenging. To keep the focus on the actual consistency relation, an extra XML-based format was established and used for the project to demonstrate the general feasibility of the approach (without actually exporting this format from the CAD tool). The MBTT scenario requires a CAM tool adapter and a parser for CNC.

In general, it is certainly complex if not impossible to incrementally update information inside of most established tools without having to resort to a typically textual export/import format. Such an exchange format and the corresponding support for it is often buggy or incomplete, can be very inefficient, or even incur unwanted information loss if extra tool-specific information such as accurate time stamps or IDs are not exported/imported.

Finally, consistency management tasks in general, and synchronisation tasks in particular work best if explicit change information in form of deltas is available and is as precise and detailed as possible [11]. In CMDAE (and in many other projects) this was not the case, however, implying that non-trivial logic must be implemented for every metamodel to be able to compare models (before and after a change), and compute the best possible delta (which is often not unique). The problem is simply that most tools (at least in the industry automation domain) do not provide access to such explicit change information. Note that using "state-based" bx approaches, i.e., providing fragments such as SYNC that take models as input and produce models as output (instead of deltas), does not help here; the delta calculation is still required and is just implemented (and hidden) as part of the bx [11].[7]

---

[7] This can have some advantages as a clean separation sometimes adds additional complexity as opposed to flexibly (e.g., recursively) mixing update propagation with change detection.

### 4.3 How do I start?

In the CMDAE scenario, the method pattern `DirectedStart` could be applied as there was a clear forward engineering process as described in section 2.1 and depicted in figure 3. Subsequent synchronisation steps could also be well supported with incremental SYNC fragments as all consistency relations are binary [37] and resolution paths are fairly straightforward. This is probably because consistency management must be performed manually today and thus cannot be arbitrarily complex. It would, however, be interesting to transition from descriptive to more prescriptive resolution networks and attempt to change and optimise the resolution process. Although this is not shown explicitly in figure 5, starting the synchronisation process in the CME project was challenging due to the requirement that the existing CNC code generator be used unchanged. This means that the initial CNC model and TP models were created by the existing tools without a means of establishing a correspondence in the process. Consequently, the method pattern `ColdStart` had to be used to establish this correspondence as required for the rest of the process. For the GraTram scenario, the method pattern `ColdStart` was clearly required as both models are developed independently and are sometimes partially reused, i.e., exist already. For the MBTT scenario, both models are created by different systems (SUT vs. bx) so the method fragments initial GEN and initial CC are required according to the pattern `GenerateAndCheck`.

In general, determining how to start the resolution process is often a crucial point (hence our method patterns) and can decide if an implemented solution is feasible in practice or not.

### 4.4 Do I need a bx language?

For the CMDAE scenario the required initial SYNC fragments can certainly be implemented in a standard programming language. The incremental SYNC fragments could also be implemented but are a bit more challenging in this respect as information loss must be avoided, i.e., the existing source/target model must be carefully updated in such a way that the resulting models are consistent but without unnecessarily discarding any information. Besides the implementation effort, the project was very iterative with constant changes to the TGG rules. This implies that a model-driven approach is certainly advantageous to avoid having to constantly update a comparably low-level implementation. At least for the CMDAE scenario, a promising approach might be to transition from a TGG to an implementation in a perhaps more scalable or otherwise preferred programming language as a final step in the project. Note that the TGG can still be used for testing purposes even after this transition [38].

For the CME scenario, scalability was an important issue due to the potentially large size of generated CNC code. Once a correspondence has been established, however, this could actually be an argument for an incremental approach that can guarantee efficient incremental updates. Starting efficiently, however, remains a challenge. Implementing the initial CC fragment for CME manually with a standard programming language is non-trivial due to the complex and non-unique correspondence between a single TP and possibly numerous consistent CNC programs.

In the GraTram scenario, the semantic overlap of simulation and CAD models is rather small compared to the remaining parts of these models. Our experience is that this makes the method fragments initial CC and incremental SYNC particularly appealing and elegant. An iterative development of the TGG is possible where one starts with just a few TGG rules and then increases the overlap as the understanding for the consistency relation grows. Similar to CME, the initial correspondence between simulation and CAD models is more an optimisation problem than a straightforward calculation [32]. While incremental SYNC can be implemented manually, it can again be difficult to carefully update existing models.

As discussed for CME, handling large CNC programs in the MBTT scenario could pose a scalability challenge. This might, however, not be so crucial for testing purposes and it might be feasible to use derived and solver-based initial CC fragments [32]. It would be challenging to implement initial CC without a bx language as the generated code is not at all unique. Multiple variation points lead to an explosion of the solution space and this is indeed the basic motivation for applying bx technology in this context.

To summarise, some method fragments are fairly straightforward to implement (initial SYNC), while others are more complex (all incremental variants and initial CC). Based on typical resolution paths for the scenario, this can already be used to decide if using a dedicated bx language such as TGGs is necessary or not. Other practical issues that affect this decision include scalability, integration in an existing tool chain that might demand a certain programming language, and how stable the current understanding of the consistency relation is. It is also important to note that a bx language can be used solely for quality assurance of certain perhaps especially crucial aspects of an implementation.

### 4.5 How formal, and on what level of abstraction should description languages be?

It is worth discussing whether description languages such as the two we have presented and used in this paper have to be at all formal. While intuitive diagrams might be sufficient as documentation, we have opted for simple but formal languages to enable future automated analyses of the project descriptions.

In the problem domain, for example, optimal resolution paths can be determined and suggested based on domain-specific objective functions. Transformation contexts can also be inspected and used to generate tests for confluence, i.e., if the final consistent transformation network depends on the specific resolution path chosen to restore consistency. Finally, an analysis could also be used to decide if a given decomposition of a consistency relation into multiple (binary) consistency relations is problematic or not (cf. Stevens' discussion [37] of this for more details). To automate (program) such tasks, a certain level of formality is necessary.

Similarly, the solution-domain description language could be used as a generic orchestration language, to be transformed to an executable program by using a collection of implemented fragments developed with or without a bx tool. To enable such an automated transformation, a certain level of formality is again required.

Concerning the chosen level of abstraction, our primary goal with the description language in the problem domain was to abstract from project-specific details to enable a comparison of all projects and an identification of common patterns. To accomplish this, we adopted Stevens' approach [37] of focussing solely on models and consistency relations between models. In the solution domain, our goal was to abstract from specific bx approaches (such as TGGs) by focussing solely on fragments and patterns describing how to combine the fragments. We believe that our goals have been achieved as demonstrated by this paper: we have been able to identify common fragments and patterns used across all projects, and our fragments and patterns can be implemented (with more or less effort) with any reasonably expressive bx tool, or of course in a general purpose programming language.

An important aspect that is still missing from the languages concerns details of available or required tooling. While this is a rather technical point, it is important if not crucial in practical scenarios. We are, for example, currently unable to capture in the project description of CME that an existing CNC code generator had to be used, even though this had an impact on applicable patterns (`ColdStart` vs. `DirectedStart`). We leave a consideration of such extensions to future work.

## 5 Related Work

In this section, we review prior work on *megamodelling, method engineering,* and *patterns* for (bidirectional) model transformations. Furthermore we reflect on existing (industrial) projects where bx methods and technologies were applied.

Our description language for the problem domain is inspired by Stevens' work on describing networks of bx [37]. Stevens considers the study of the resolution of networks of bx as a contributing technology for megamodelling [3] and discusses the task of consistency restoration of a network of connected models. In this paper we adapt and extend Stevens' description language by (i) introducing the concept of a transformation schema to control the general shape of valid networks, and (ii) adding swim lanes in the visual notation to indicate domains and improve readability. We also demonstrate that the language and notation can be used to describe projects of realistic complexity.

Our description language for the solution domain comes from the research area of method engineering [5]. The main purpose of a *method* is to guide a complex software engineering task, such as the development of a software system, its transformation/migration, or in our case the consistency management of involved abstractions of the same system. A method supports and describes this task by specifying the activities to enact, artefacts to generate, tools to use or roles to involve [13]. A specific manifestation of method engineering is *Situational Method Engineering* (SME) which encompasses all aspects of creating a method for a specific situation [22]. Approaches that follow the SME paradigm consider the situational context in which a method will be applied during the development of the method, so that it can be adapted to the context and is then called situation-specific. In the context of SME, reusable, atomic building blocks of a method, e.g., a single activity, artefact, role or tool, are called method fragments. To increase the efficiency of method development, *method patterns* can be used to provide additional guidance concerning how to combine method fragments. A pattern in general is associated with a reoccurring problem in a certain context [1], for which it describes the core of a solution. Method patterns transfer this idea to the field of method engineering [16].

The active research area of megamodelling (we refer to Härtel et al. [21] for a recent survey of megamodelling approaches) provides other viable alternatives to Stevens' networks of bx [37], and Engels' method engineering approach [13]. We discuss a few of these megamodelling languages in the following.

*MegaL* [15] is a megamodelling language that could be used as a basis for both our problem and solution domain description languages. An advantage of *MegaL* is that it would be easy to check for wellformedness and formalise rigorously with adequate tool support, e.g., the connection between our two languages, or a transformation of problem domain to solution domain descriptions. *MegaL* is, however, also more general and does not support any direct concepts for consistency management. Lämmel has applied *MegaL* to the domain of *coupled transformations* (cx) to realise *LAL* [28], a megamodelling language particularly suitable for describing consistency management scenarios. Lämmel has identified various basic cx patterns such as state-based or delta-based strategies, and formalised these patterns in *LAL*. As a basis for our languages, *LAL* would enable a more rigorous handling of compatibility conditions between fragments, enabling, e.g., automated testing of implemented fragments to ensure that they do not contradict each other. While *LAL* would be a more suitable basis for our languages than *MegaL*, it still has quite a broad scope and does not distinguish between problem or solutions domains, or between fragments and patterns. For our work a suitable concrete syntax (swim lanes, story board representation of resolution paths) is also very important. Compiling our proposed languages to *LAL* programs is, however, not ruled out by our current choice of basing our languages on Stevens' work [37] and method engineering [13], and we leave this together with an exploration of potential benefits, to future work.

There has been considerable work on classifying and formalising both bx problems and solutions. A three-dimensional taxonomy for model synchronisation scenarios is presented by Diskin et al. [10] covering different scenario types with respective requirements and properties. Further results in a similar direction are provided by Hidaka et al. [25], classifying bx approaches using a feature model, Eramo et al. [14], illustrating a set of relevant properties pertaining to bidirectional model transformations, and Lano et al. [29] proposing patterns for specifying bidirectional transformations with their tool UML-RSDS. Finally, Diskin et al. [9] formalise synchronisation operations by providing a suitable algebraic framework. Compared to these results, our paper covers *both* the problem domain and an important solution domain for consistency management. We take a broader view on consistency management, focussing not only on synchronisation but covering also model generation and consistency checking. Our contribution in this paper is to propose languages for *describing* scenarios and solution strategies in a tool-independent manner, leaving a formalisation of desired/guaranteed properties or compatibility conditions between fragments to future work.

Concerning our example scenarios from industry automation; bx approaches have been applied in various other industrial domains: Hermann et al. [23] report on using TGGs to translate satellite procedures, Blouin et al. [4] demonstrate an incremental synchronisation layer between textual and graphical editors, Giese et al. [17] present a synchronisation solution for SysML and AUTOSAR models, while Cunha et al. [7] propose a bidirectional transformation approach for model-driven spreadsheets. Using our description languages, such projects could provide their results, solution strategies, and lessons learnt in a uniform and comparable manner; this would enable the identification and reuse of further method patterns for consistency management.

## 6 Conclusion and Future Work

In this paper, we have proposed description languages for consistency management scenarios both in the problem domain and in an MDE solution domain. We demonstrated the benefit of our languages by describing a series of projects we investigated in the industry automation domain together with our industrial partner Siemens AG. Table 1 provides a tabular summary of the discussion in section 4.

Our description languages, especially in the solution domain, are heavily influenced by our MDE and TGG outlook and experience. For instance, our method fragments and method patterns are uniformly *delta-based*, taking "arrows" as input and output instead of models. We believe this provides for conceptual clarity as it generalises the state-based case and separates delta discovery from update propagation. In this respect and in many others, our suggestions are not meant as a perfect version but rather as a good start for a community discussion and consensus on general description languages for bx. Just as we have adapted and extended Stevens' suggestions for the problem domain, we welcome further collaboration and discussion from other diverse perspectives on bx.

In the research projects described in this paper, we mostly concentrated on what was relevant and interesting from a research point of view. For instance, implementing complete and industrial strength parsers and model diffs for delta discovery was clearly out of scope. We believe, however, that such practical challenges can be addressed. In particular, a crucial point is to utilise suitable abstractions provided by engineering tools and avoid operating directly on engineering data represented at lower levels of abstraction. From a tooling point of view, a further point is that engineering tools should allow for an *incremental* manipulation of their contained data based on these provided abstractions.

**Table 1** Consistency Management Projects in the Industry Automation Domain

| Project | Involved Roles | Domain | Primary Challenge | Patterns |
| --- | --- | --- | --- | --- |
| CMDAE | Electrical Engineer, Automation Engineer, Software Engineer | Automation Engineering | Engineering of mutually dependent models | DirectedStart |
| CME | Product Designer, Manufacturing Engineer, Machine Operator | Manufacturing Engineering | Integration of an existing CNC code generator | ColdStart |
| GraTram | CAD Engineer, Simulation Engineer | Computer Aided Engineering | Starting with independently developed models | ColdStart |
| MBTT | Test Engineer | Quality Assurance in Manufacturing Engineering | Testing a non-deterministic code generator | GenerateAndCheck |

The resolution paths described in this paper are descriptive and not prescriptive. This means that we aimed to support existing workflows typically by automating certain steps. As Stevens [37] points out, the next step would be to think about *optimising* resolution paths and even consistency relations. Indeed with appropriate automation, some resolution paths that were previously infeasible to execute manually might become attractive. Further research is required to identify and analyse various desirable properties of networks of bx, and to provide corresponding optimisations.

Analogously, but in the solution domain, a further task is to identify, formalise, and guarantee desirable properties of our method fragments. Method patterns should also preserve (some of) these properties for their required composition of fragments.

Finally, we restricted our presentation of method patterns to the main patterns that were repeatedly used for the research projects described in this paper. Further (catalogues of) consistency management patterns include (i) different strategies for handling concurrent changes in multiple domains [39], (ii) strategies for handling traceability, mapping, change detection, and (resource) allocation scenarios involving a synergistic combination of bx approaches (e.g., TGGs and constraints solvers [32]), and (iii) strategies for coping with non-deterministic scenarios [2] requiring (a combination of) user interaction and an optimisation phase.

## References

[1] Christopher Alexander, Sara Ishikawa, Murray Silverstein, Max Jacobson, Ingrid Fiksdahl-King, and Shlomo Angel. *A Pattern Language - Towns, Buildings, Construction*. Oxford University Press, 1977. ISBN: 978-0-19-501919-3.

[2] Anthony Anjorin, Erhan Leblebici, Andy Schürr, and Gabriele Taentzer. "A Static Analysis of Non-Confluent Triple Graph Grammars for Efficient Model Transformation". In: *Graph Transformation - 7th International Conference, ICGT 2014, Held as Part of STAF 2014, York, UK, July 22-24, 2014. Proceedings*. Volume 8571. LNCS. Springer, 2014, pages 130–145. DOI: 10.1007/978-3-319-09108-2_9.

[3] Jean Bézivin, Frédéric Jouault, and Patrick Valduriez. "On the Need for Megamodels". In: *Proceedings of the OOPSLA/GPCE: Best Practices for Model-Driven Software Development workshop, 19th Annual ACM Conference on Object-Oriented Programming, Systems, Languages, and Applications*. 2004.

[4] Dominique Blouin, Alain Plantec, Pierre Dissaux, Frank Singhoff, and Jean-Philippe Diguet. "Synchronization of Models of Rich Languages with Triple Graph Grammars: An Experience Report". In: *Theory and Practice of Model Transformations - 7th International Conference, ICMT 2014, Held as Part of STAF 2014, York, UK, July 21-22, 2014. Proceedings*. Edited by Davide Di Ruscio and Dániel Varró. Volume 8568. LNCS. Springer, 2014, pages 106–121. DOI: 10.1007/978-3-319-08789-4_8.

[5]   Sjaak Brinkkemper. "Method Engineering: Engineering of Information Systems Development Methods and Tools". In: *Information and Software Technology* 38.4 (1996), pages 275–280. DOI: 10.1016/0950-5849(95)01059-9.

[6]   Antonio Cicchetti, Davide Di Ruscio, Romina Eramo, and Alfonso Pierantonio. "JTL : A Bidirectional and Change Propagating Transformation Language". In: *Software Language Engineering - Third International Conference, SLE 2010, Eindhoven, The Netherlands, October 12-13, 2010, Revised Selected Papers*. Edited by Brian A. Malloy, Steffen Staab, and Mark van den Brand. Volume 6563. LNCS. Springer, 2010, pages 183–202. DOI: 10.1007/978-3-642-19440-5_11.

[7]   Jácome Cunha, João P. Fernandes, Jorge Mendes, Hugo Pacheco, and João Saraiva. "Bidirectional Transformation of Model-Driven Spreadsheets". In: *Theory and Practice of Model Transformations - 5th International Conference, ICMT 2012, Prague, Czech Republic, May 28-29, 2012. Proceedings*. Edited by Zhenjiang Hu and Juan de Lara. Volume 7307. LNCS. Springer, 2012, pages 105–120. DOI: 10.1007/978-3-642-30476-7_7.

[8]   Krzysztof Czarnecki, John Nathan Foster, Zhenjiang Hu, Ralf Lämmel, Andy Schürr, and James Terwilliger. "Bidirectional Transformations: A Cross-Discipline Perspective". In: *Theory and Practice of Model Transformations, Second International Conference, ICMT 2009, Zurich, Switzerland, June 29-30, 2009. Proceedings*. Edited by Richard F. Paige. Volume 5563. LNCS. Springer, 2009, pages 260–283. DOI: 10.1007/978-3-642-02408-5_19.

[9]   Zinovy Diskin. "Algebraic Models for Bidirectional Model Synchronization". In: *Model Driven Engineering Languages and Systems, 11th International Conference, MoDELS 2008, Toulouse, France, September 28 - October 3, 2008. Proceedings*. Edited by Krzysztof Czarnecki, Ileana Ober, Jean-Michel Bruel, Axel Uhl, and Markus Völter. Volume 5301. LNCS. Springer, 2008, pages 21–36. DOI: 10.1007/978-3-540-87875-9_2.

[10]  Zinovy Diskin, Hamid Gholizadeh, Arif Wider, and Krzysztof Czarnecki. "A Three-dimensional Taxonomy for Bidirectional Model Synchronization". In: *Journal of Systems and Software* 111 (2016), pages 298–322. DOI: 10.1016/j.jss.2015.06.003.

[11]  Zinovy Diskin, Yingfei Xiong, Krzysztof Czarnecki, Hartmut Ehrig, Frank Hermann, and Fernando Orejas. "From State- to Delta-Based Bidirectional Model Transformations: the Symmetric Case". In: *Model Driven Engineering Languages and Systems, 14th International Conference, MODELS 2011, Wellington, New Zealand, October 16-21, 2011. Proceedings*. Edited by Jon Whittle, Tony Clark, and Thomas Kühne. Volume 6981. LNCS. Springer, 2011, pages 304–318. DOI: 10.1007/978-3-642-24485-8_22.

[12]  Hartmut Ehrig, Karsten Ehrig, Ulrike Prange, and Gabriele Taentzer. *Fundamentals of Algebraic Graph Transformation*. Edited by Wilfried Brauer, Grzegorz Rozenberg, and Arto Salomaa. Monographs in Theoretical Computer Science. An EATCS Series. Springer, 2006. ISBN: 3-540-31187-4. DOI: 10.1007/3-540-31188-2.

[13] Gregor Engels and Stefan Sauer. "A Meta-Method for Defining Software Engineering Methods". In: *Graph Transformations and Model-Driven Engineering - Essays Dedicated to Manfred Nagl on the Occasion of his 65th Birthday*. Edited by Gregor Engels, Claus Lewerentz, Wilhelm Schäfer, Andy Schürr, and Bernhard Westfechtel. Volume 5765. LNCS. Springer, 2010, pages 411–440. DOI: 10.1007/978-3-642-17322-6_18.

[14] Romina Eramo, Romeo Marinelli, and Alfonso Pierantonio. "Towards a Taxonomy for Bidirectional Transformation". In: *Post-proceedings of the Seventh Seminar on Advanced Techniques and Tools for Software Evolution, SATToSE 2014, L'Aquila, Italy, 9-11 July 2014*. Edited by Davide Di Ruscio and Vadim Zaytsev. Volume 1354. CEUR Workshop Proceedings. CEUR-WS.org, 2014, pages 122–131.

[15] Jean-Marie Favre, Ralf Lämmel, and Andrei Varanovich. "Modeling the Linguistic Architecture of Software Products". In: *Model Driven Engineering Languages and Systems - 15th International Conference, MODELS 2012, Innsbruck, Austria, September 30-October 5, 2012. Proceedings*. Edited by Robert B France, Jürgen Kazmeier, Ruth Breu, and Colin Atkinson. Volume 7590. LNCS. Springer, 2012, pages 151–167. DOI: 10.1007/978-3-642-33666-9_11.

[16] Masud Fazal-Baqaie, Markus Luckey, and Gregor Engels. "Assembly-Based Method Engineering with Method Patterns". In: *Software Engineering 2013 - Workshopband (inkl. Doktorandensymposium), Fachtagung des GI-Fachbereichs Softwaretechnik*. Edited by Stefan Wagner and Horst Lichter. Volume 215. LNI. GI, 2013, pages 435–444.

[17] Holger Giese, Stephan Hildebrandt, and Stefan Neumann. "Model Synchronization at Work : Keeping SysML and AUTOSAR Models Consistent". In: *Graph Transformations and Model-Driven Engineering - Essays Dedicated to Manfred Nagl on the Occasion of his 65th Birthday*. Edited by Gregor Engels, Claus Lewerentz, Wilhelm Schäfer, Andy Schürr, and Bernhard Westfechtel. Volume 5765. LNCS. Springer, 2010, pages 555–579. DOI: 10.1007/978-3-642-17322-6_24.

[18] Susann Gottmann, Frank Hermann, Claudia Ermel, Thomas Engel, and Gianluigi Morelli. "Towards Bidirectional Engineering of Satellite Control Procedures using Triple Graph Grammars". In: *Proceedings of the 7th Workshop on Multi-Paradigm Modeling co-located with the 16th International Conference on Model Driven Engineering Languages and Systems, MPM@MoDELS 2013, Miami, Florida, September 30, 2013*. Edited by Christophe Jacquet, Daniel Balasubramanian, Edward Jones, and Tamás Mészáros. Volume 1112. CEUR Workshop Proceedings. CEUR-WS.org, 2013, pages 67–76.

[19] Marvin Grieger and Masud Fazal-baqaie. "Towards a Framework for the Modular Construction of Situation- Specific Software Transformation Methods". In: *Softwaretechnik-Trends* 35.2 (2015).

[20] Marvin Grieger, Masud Fazal-baqaie, Gregor Engels, and Markus Klenke. "Concept-Based Engineering of Situation-Specific Migration Methods". In: *Software Reuse: Bridging with Social-Awareness - 15th International Conference, ICSR 2016, Limassol, Cyprus, June 5-7, 2016, Proceedings*. Edited by M. Georgia; Kapitsaki and

Eduardo Santana de Almeida. Volume 9679. LNCS. Springer, 2016, pages 199–214. DOI: 10.1007/978-3-319-35122-3_14.

[21] Johannes Härtel, Lukas Härtel, Ralf Lämmel, Andrei Varanovich, and Marcel Heinz. "Interconnected Linguistic Architecture". In: *Programming Journal* 1.1 (2017), 3:1–3:27. DOI: 10.22152/programming-journal.org/2017/1/3.

[22] Brian Henderson-Sellers, Jolita Ralyté, Pär J. Ågerfalk, and Matti Rossi. *Situational Method Engineering*. Springer, 2014. ISBN: 978-3-642-41466-4. DOI: 10.1007/978-3-642-41467-1.

[23] Frank Hermann, Susann Gottmann, Nico Nachtigall, Hartmut Ehrig, Benjamin Braatz, Gianluigi Morelli, Alain Pierre, Thomas Engel, and Claudia Ermel. "Triple Graph Grammars in the Large for Translating Satellite Procedures". In: *Theory and Practice of Model Transformations - 7th International Conference, ICMT 2014, Held as Part of STAF 2014, York, UK, July 21-22, 2014. Proceedings*. Edited by Davide Di Ruscio and Dániel Varró. Volume 8568. LNCS. Springer, 2014, pages 122–137. DOI: 10.1007/978-3-319-08789-4_9.

[24] Soichiro Hidaka, Zhenjiang Hu, Kazuhiro Inaba, Hiroyuki Kato, and Keisuke Nakano. "GRoundTram: An Integrated Framework for Developing Well-Behaved Bidirectional Model Transformations". In: *26th IEEE/ACM International Conference on Automated Software Engineering (ASE 2011), Lawrence, KS, USA, November 6-10, 2011*. Edited by Perry Alexander, Corina S. Pasarenau, and John G. Hosking. IEEE Computer Society, 2011, pages 480–483. DOI: 10.1109/ASE.2011.6100104.

[25] Soichiro Hidaka, Massimo Tisi, Jordi Cabot, and Zhenjiang Hu. "Feature-Based Classification of Bidirectional Transformation Approaches". In: *Software and System Modeling* 15.3 (2016), pages 907–928. DOI: 10.1007/s10270-014-0450-0.

[26] Stephan Hildebrandt, Leen Lambers, Holger Giese, Dominic Petrick, and Ingo Richter. "Automatic Conformance Testing of Optimized Triple Graph Grammar Implementations". In: *Applications of Graph Transformations with Industrial Relevance - 4th International Symposium, AGTIVE 2011, Budapest, Hungary, October 4-7, 2011, Revised Selected and Invited Papers*. Edited by Andy Schürr, Dániel Varró, and Gergely Varró. Volume 7233. LNCS. Springer, 2011, pages 238–253. DOI: 10.1007/978-3-642-34176-2_20.

[27] Hsiang-shang Ko, Tao Zan, and Zhenjiang Hu. "BiGUL: A Formally Verified Core Language for Putback-Based Bidirectional Programming". In: *Proceedings of the 2016 ACM SIGPLAN Workshop on Partial Evaluation and Program Manipulation, PEPM 2016, St. Petersburg, FL, USA, January 20 - 22, 2016*. Edited by Martin Erwig and Tiark Rompf. ACM, 2016, pages 61–72. DOI: 10.1145/2847538.2847544.

[28] Ralf Lämmel. "Coupled Software Transformations Revisited". In: *Proceedings of the 2016 ACM SIGPLAN International Conference on Software Language Engineering, Amsterdam, The Netherlands, October 31 - November 1, 2016*. Edited by Tijs van der Storm, Emilie Balland, and Dániel Varró. ACM, 2016, pages 239–252. DOI: 10.1145/2997364.

[29]  Kevin Charles Lano, Hessa Alfraihi, Sobhan Yassipour Tehrani, and Howard Haughton. "Patterns for Specifying Bidirectional Transformations in UML-RSDS". In: *The 10th International Conference on Software Engineering Advances (ICSEA 2015)*. IARIA XPS Press, 2015.

[30]  Marius Lauder. "Incremental Model Synchronization with Precedence-Driven Triple Graph Grammars". PhD thesis. Technische Universität Darmstadt, 2012.

[31]  Erhan Leblebici, Anthony Anjorin, and Andy Schürr. "Developing eMoflon with eMoflon". In: *Theory and Practice of Model Transformations - 7th International Conference, ICMT 2014, Held as Part of STAF 2014, York, UK, July 21-22, 2014. Proceedings*. Volume 8568. LNCS. Springer, 2014, pages 138–145. DOI: 10.1007/978-3-319-08789-4_10.

[32]  Erhan Leblebici, Anthony Anjorin, and Andy Schürr. "Inter-model Consistency Checking using Triple Graph Grammars and Linear Optimization Techniques". In: *Fundamental Approaches to Software Engineering - 20th International Conference, FASE 2017, Held as Part of the European Joint Conferences on Theory and Practice of Software, ETAPS 2017, Uppsala, Sweden, April 22-29, 2017, Proceedings*. Edited by Marieke Huisman and Julia Rubin. Volume 10202. LNCS. Springer, 2017, pages 191–207. DOI: 10.1007/978-3-662-54494-5_11.

[33]  Nuno Macedo, Tiago Guimarães, and Alcino Cunha. "Model Repair and Transformation with Echo". In: *2013 28th IEEE/ACM International Conference on Automated Software Engineering, ASE 2013, Silicon Valley, CA, USA, November 11-15, 2013*. Edited by Ewen Denney, Tevfik Bultan, and Andreas Zeller. IEEE, 2013, pages 694–697. DOI: 10.1109/ASE.2013.6693135.

[34]  Sebastian Rose, Marius Lauder, Michael Schlereth, and Andy Schürr. "A Multi-dimensional Approach for Concurrent Model Driven Automation Engineering". In: *Model-Driven Domain Analysis and Software Development: Architectures and Functions*. Edited by Janis Osis and Erika Asnina. IGI Publishing, 2011, pages 90–113. DOI: 10.4018/978-1-61692-874-2.ch005.

[35]  Andy Schürr. "Specification of Graph Translators with Triple Graph Grammars". In: *Graph-Theoretic Concepts in Computer Science, 20th International Workshop, WG '94, Herrsching, Germany, June 16-18, 1994, Proceedings*. Edited by Ernst W. Mayr, Gunther Schmidt, and Gottfried Tinhofer. Volume 903. LNCS. Springer, 1994, pages 151–163. DOI: 10.1007/3-540-59071-4_45.

[36]  Perdita Stevens. "A Landscape of Bidirectional Model Transformations". In: *Generative and Transformational Techniques in Software Engineering II: International Summer School, GTTSE 2007, Braga, Portugal, July 2-7, 2007. Revised Papers*. Edited by Ralf Lämmel, Joost Visser, and João Saraiva. Volume 5235. LNCS. Springer, 2008, pages 408–424. DOI: 10.1007/978-3-540-88643-3_10.

[37]  Perdita Stevens. "Bidirectional Transformations in the Large". In: *20th ACM/IEEE International Conference on Model Driven Engineering Languages and Systems, MODELS 2017, Austin, TX, USA, September 17-22, 2017*. IEEE Computer Society, 2017. DOI: 10.1109/MODELS.2017.8.

[38]   Martin Wieber, Anthony Anjorin, and Andy Schürr. "On the Usage of TGGs for Automated Model Transformation Testing". In: *Theory and Practice of Model Transformations - 7th International Conference, ICMT 2014, Held as Part of STAF 2014, York, UK, July 21-22, 2014. Proceedings*. Edited by Davide Di Ruscio and Dániel Varró. Volume 8568. LNCS. Springer, 2014, pages 1–16. DOI: 10.1007/978-3-319-08789-4_1.

[39]   Yingfei Xiong, Hui Song, Zhenjiang Hu, and Masato Takeichi. "Synchronizing Concurrent Model Updates Based on Bidirectional Transformation". In: *Software and Systems Modeling* 12.1 (2013), pages 89–104. DOI: 10.1007/s10270-010-0187-3.

**About the authors**

**Anthony Anjorin** is a junior professor for model-based software development at Paderborn University. Contact him at anthony.anjorin@upb.de.

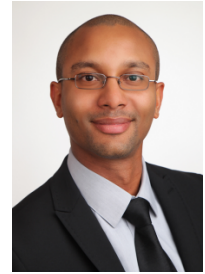

**Enes Yigitbas** is a PhD student at Paderborn University. Contact him at enes@mail.uni-paderborn.de.

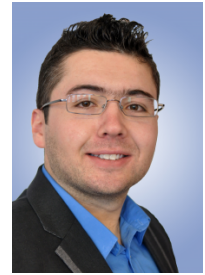

**Erhan Leblebici** is a PhD student at Technische Universität Darmstadt. Contact him at erhan.leblebici@es.tu-darmstadt.de.

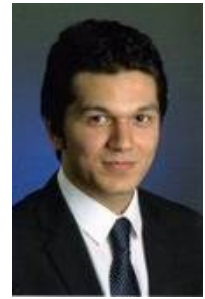

**Andy Schürr** is a full professor at Technische Universität Darmstadt. Contact him at andy.schuerr@es.tu-darmstadt.de.

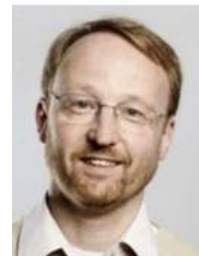

**Marius Lauder** is head of embedded HMI-framework development at the German car supplier Continental. Contact him at marius.lauder@continental-corporation.com.

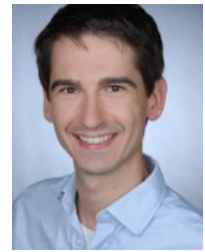

**Description Languages for Consistency Management Scenarios**

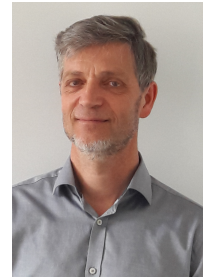

**Martin Witte** is Senior Principal Expert for System Engineering and Simulation at Siemens AG in Nuremberg. Contact him at martin.witte@siemens.com.